\documentclass{aa}
\usepackage{graphicx,fixltx2e,hyperref}
\usepackage{txfonts,color}

\usepackage{natbib} 
\bibpunct{(}{)}{;}{a}{}{,}

\begin{document}

 \title{Can spicules be detected at disc centre in broad-band \ion{Ca}{ii} H filter imaging data ?}

   \author{C. Beck\inst{1,2,3} \and R. Rezaei\inst{4} \and K.G. Puschmann\inst{5}}
        
   \titlerunning{Spicules at disc centre in \ion{Ca}{ii} H filter imaging data ?}
  \authorrunning{C. Beck, R. Rezaei, K.G. Puschmann}  
\institute{Instituto de Astrof\'isica de Canarias (IAC)
     \and Departamento de Astrof\'isica, Universidad de La Laguna
     \and National Solar Observatory (NSO)
     \and Kiepenheuer-Institut f\"ur Sonnenphysik (KIS)
       \and Leibniz-Institut f\"ur Astrophysik Potsdam (AIP)
    }
 
\date{Received xxx; accepted xxx}

\abstract{Recently, a possible identification of type-II spicules in
  broad-band (full-width at half-maximum (FWHM) of $\sim$0.3\,nm)
  filter imaging data in \ion{Ca}{ii} H on the solar disc was reported.}{We
  estimate the formation height range contributing to broad-band and
    narrow-band filter imaging data in \ion{Ca}{ii} H to investigate
    whether spicules can be detected in such observations at the 
    centre of the solar disc.}{We apply spectral filters of FWHMs from 0.03\,nm to
    1\,nm to observed \ion{Ca}{ii} H line profiles to simulate Ca imaging
    data. We use observations across the limb to estimate the relative
  intensity contributions of off-limb and on-disc structures. We compare the
  synthetic Ca filter imaging data with intensity maps of Ca spectra at
    different wavelengths and temperature maps at different optical depths
    obtained by an inversion of these spectra. In addition, we
  determine the intensity response function for the wavelengths
  covered by the filters of different FWHM.}{In broad-band 
    (FWHM\,=\,0.3\,nm) {\bf Ca\rm} imaging data, the intensity emitted off
  the solar limb is about 5\,\% of the intensity at disc
  centre.  For a 0.3\,nm-wide filter centred at the \ion{Ca}{ii} H line
    core, up to about one third of the off-limb intensity comes from emission
    in H$\epsilon$. On the disc, only about 10 to 15\,\% of the intensity transmitted through
  a broad-band filter comes from the line-core region between the H$_1$ minima
  (396.824 to 396.874\,nm). No traces of elongated fibrillar structures
  are visible in the synthetic Ca broad-band imaging data at disc centre\rm, opposite to the line-core images of the Ca spectra. The intensity-weighted response function
  for a 0.3\,nm-wide filter centred at the \ion{Ca}{ii} H line core peaks at
  about $\log \tau \sim -2$ ($z \sim 200 $\,km\rm). Relative contributions
  from atmospheric layers above 800\,km are about 10\,\%. The inversion
  results suggest that the slightly enhanced emission around the photospheric
  magnetic network  in broad-band Ca imaging data  is caused by a thermal canopy at a height of about 600\,km.}{Broad-band ($\sim$0.3\,nm) \ion{Ca}{ii} H imaging data do not trace upper chromospheric structures such as spicules in observations at  the solar disc because of the too small relative contribution of the line core to the total wavelength-integrated filter intensity. The faint haze around network elements in broad-band Ca imaging observations at disc centre presumably traces thermal canopies in the vicinity of magnetic flux concentrations instead.} 
\maketitle
\section{Introduction}
One of the most successful recent broad-band ($\sim$0.3\,nm) filter-imaging systems in use in solar physics is the imaging channel in the chromospheric \ion{Ca}{ii} H line on-board the Hinode satellite \citep{kosugi+etal2007,tsuneta+etal2008}. The seeing-free data from the Hinode mission are an unique source of high-quality solar observations at a constant spatial resolution. One topic that was basically revived by the advent of the Hinode Ca imaging data were solar spicules that are seen in chromospheric lines at and beyond the solar limb \citep{beckers1968,sterling00}. The Hinode Ca imaging data triggered a whole series of investigations because for the first time the temporal evolution of individual spicules could be followed in detail \citep[cf.][]{depontieu+etal2007,sterling+etal2010}.

On-disc counterparts of spicules were identified in observed line spectra by
their velocity signature, i.e.~rapid Doppler excursions of chromospheric lines
towards the blue that imply fast upwards motions  \citep[][but see also
\citet{judge+etal2011}]{langangen+etal2008,rouppevandervoort2009,sekse+etal2012}.
Recently, \citet[][DW12]{dewijn2012} interpreted a faint intensity haze around
the photospheric magnetic network in a difference image of Hinode Ca
  imaging data and simultaneous line-core intensity maps of the photospheric
  \ion{Fe}{i} line at 630.15\,nm as the signature of type-II spicules. He argued that the Hinode Ca filter has some significant chromospheric contribution that would produce this haze as a consequence of chromospheric features. Whereas this argument holds for Lyot-type filters \citep[e.g.][]{oehman1938,lyot1944,wang+etal1995,skomorovsky+etal2012} that can have spectral pass-bands down to 0.01\,nm \citep[e.g.][RE09]{reardon+etal2009}, it is not immediately clear whether the same is true for broad-band interference filters with 0.1 to 1\,nm full-width at half-maximum (FWHM) band-passes. 

The investigation of the intensity response of the \ion{Ca}{ii} H broad-band
filter of Hinode by \citet{carlsson+etal2007} yielded a maximal response at a
height of about 250\,km, whereas \citet[][]{jafarzadeh+etal2013} reported
about 450\,km for the 0.18\,nm filter of SuFi on-board of the Sunrise balloon
mission \citep{barthol+etal2011}. \citet{pietarila+etal2009} found that 90\,\%
of the intensity in a 0.15\,nm wide Ca filter should originate from layers
below 500\,km. RE09 found ``{\em no significant chromospheric signature in the
  Hinode/SOT Ca II H quiet-Sun filtergrams}'' from a comparison to
\ion{Ca}{II} IR spectra. An identification of type-II spicules in broad-band
Ca imaging data at disc centre is thus rather surprising, but we note that
  the results above referred to the quiet Sun with low magnetic activity.
 
Outside the solar limb even broad-band Ca filter imaging data trace without
doubt chromospheric structures because of the absence of any photospheric
radiation, but it is not clear if the same applies to observations on the
solar disc. Here, we investigate the contribution of off-limb features
  such as spicules to the wavelength-integrated intensity for the case of
broad-band filter observations on the solar disc by comparing the latter
  with resolved \ion{Ca}{ii} H spectra and synthetic imaging data obtained by
  a multiplication of the spectra with filters of different FWHM.

Section \ref{obs} describes the various data sets used. The observational
results of the \ion{Ca}{ii} H spectroscopy and the (synthetic) imaging data at
the centre of the solar disc and at the solar limb are presented in
Sect.~\ref{results}, together with the calculation of theoretical intensity
response functions. The results are discussed in Sect.~\ref{discus}, whereas
Sect.~\ref{schluss} provides our conclusions. Appendix \ref{hepsi}
  investigates the relative contribution of the chromospheric H$\epsilon$ line
  at 397\,nm to the Hinode Ca broad-band prefilter.
\section{Observations and creation of synthetic Ca images\label{obs}}
\begin{figure}
\resizebox{8.8cm}{!}{\includegraphics{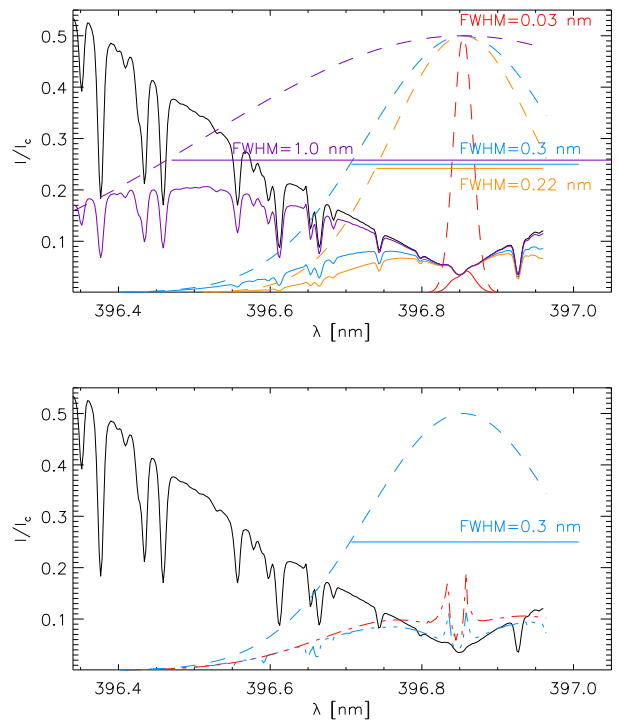}}
\caption{Average observed and theoretical Ca profiles with different filters. {\em Top panel}: average observed Ca profile at disc centre ({\em black line}). {\em Purple/blue/orange/red dashed lines}: Gaussians with FWHM of 1\,nm, 0.3\,nm, 0.22\,nm, and 0.03\,nm, respectively, centred at 396.85\,nm. {\em Purple/blue/orange/red solid lines}: multiplication of the observed profile with the filter transmission. {\em Bottom panel}: multiplication of NLTE FALC/FALF profile ({\em blue dotted/red dash-dotted lines}) profile with a 0.3\,nm filter.\label{filter}}
\end{figure}

The primary data used in this study are two observations of \ion{Ca}{ii} H spectra obtained with the POlarimetric LIttrow Spectrograph \citep[POLIS;][]{beck+etal2005b}. One data set contains Ca spectra at and
beyond the solar limb, and is described in detail in \citet{beck+rezaei2011}
and \citet{beck+etal2011}. These data are also available on-line
  \citep{beck+rezaei2011ff}. The second POLIS data set was taken at the
centre of the solar disc and is described in detail in
\citeauthor{beck+etal2009} (\citeyear{beck+etal2009}, BE09; \citeyear{beck+etal2012}). Figures \ref{limb_fov} and
\ref{center_fov} below show overview maps of these observations. Because the spectral range provided by the default POLIS Ca CCD
  (cf.~Fig.~\ref{filter}) does not cover the FWHM of 1\,nm required for the
  broadest Ca filter under investigation, we used a data set recorded at disc centre on 29 Jun 2010 with a PCO 4000 as Ca camera inside of POLIS instead \citep[cf.~the setup described in][]{beck+rezaei2012}.

For the comparison to the results derived from the spectra, we used images in
\ion{Ca}{ii} K taken with a Lyot filter of a FWHM of 0.03\,nm. This Lyot
filter is part of the slit-jaw (SJ) camera system \citep{kentischer1995} at the German Vacuum Tower Telescope \citep[VTT;][]{schroeter+soltau+wiehr1985}. In case of the disc
centre observations taken on 21 Aug 2006, the Lyot filter was mounted in front
of the Triple Etalon SOlar Spectrometer
\citep[TESOS;][]{kentischer+etal1998,tritschler+etal2002} as part of an imaging
setup \citep[cf.][]{beck+etal2007aa}, whereas near the limb (24 Apr 2011), a
PCO 4000 was mounted as SJ camera instead of the default video camera (see
\citeauthor{marian+etal2012} \citeyear{marian+etal2012} for the science data
of this observation). We also use two \ion{Ca}{ii} H images at
  similar locations from the Hinode Solar Optical Telescope
\citep{kosugi+etal2007,tsuneta+etal2008}. They were taken on 27 Aug 2011 and on 19 Feb 2007 near disc centre and at the limb,
respectively. A third Ca image at disc centre was recorded with a 1-nm wide
pre-filter that was mounted in an additional imaging channel in front of the
G\"ottingen Fabry-Perot Interferometer
\citep[GFPI;][]{puschmann+etal2006,nazi+kneer2008a} for complementary
ground-based observations during the SUNRISE flight in 2009. The \ion{Ca}{ii}
H imaging data used here were observed on 07 June 2009 in combination with
G-band imaging and spectropolarimetric measurements with the GFPI in the
\ion{Fe}{i} line at 630.25\,nm. Details on these data and the applied image
reconstruction techniques, as for example the (multi-object) multi-frame blind deconvolution \citep[\mbox{[MO]}MFBD;][]{vannortetal05} and speckle reconstruction \citep{puschmann+sailer2006}, can be found in \citet{puschmann+beck2011}.

To simulate the Hinode Ca imaging data, we multiplied all spectrally resolved observed line profiles with Gaussians of FWHM\,=\,0.22\,nm \citep{carlsson+etal2007} and 0.3\,nm \citep{tsuneta+etal2008}, respectively (cf.~Fig.~\ref{filter}, called ``broad-band'' imaging in the following). For the simulation of images taken with a Lyot-type filter, we multiplied the spectra observed with the default POLIS Ca CCD with a transmission profile of FWHM\,=\,0.03\,nm (called ``narrow-band'' imaging in the following). Finally, for the comparison with the filter images of FWHM\,=\,1\,nm, we multiplied the Ca spectra observed with the PCO 4000 camera inside of POLIS with the transmission profile of the corresponding filter.

Applying the respective filters to the average spectrum observed at disc centre ({\em top panel} of Fig.~\ref{filter}) reveals that in case of the narrow-band filter, only wavelengths at and near the line core are sampled, with the maximum of the transmitted intensity at the line core. Contrary to that, the maximum of the transmitted intensity is found far away from the line core or even the $H_1$ intensity minima for all broad-band filters. The spectral wavelength range, in which the transmitted intensity in the line wing is larger than in the line core, is also significantly wider than the line-core region itself for broad-band imaging. In the following, the synthetic broad-band data corresponding to the filter with a FWHM of 0.3\,nm will be used as being representative for the Hinode Ca imaging data, because the differences between the two broad-band filters of 0.22\,nm and 0.3\,nm FWHM are minor (cf.~Fig.~\ref{contrib1} below).
\begin{figure}
\resizebox{8.8cm}{!}{\hspace*{1.5cm}\includegraphics{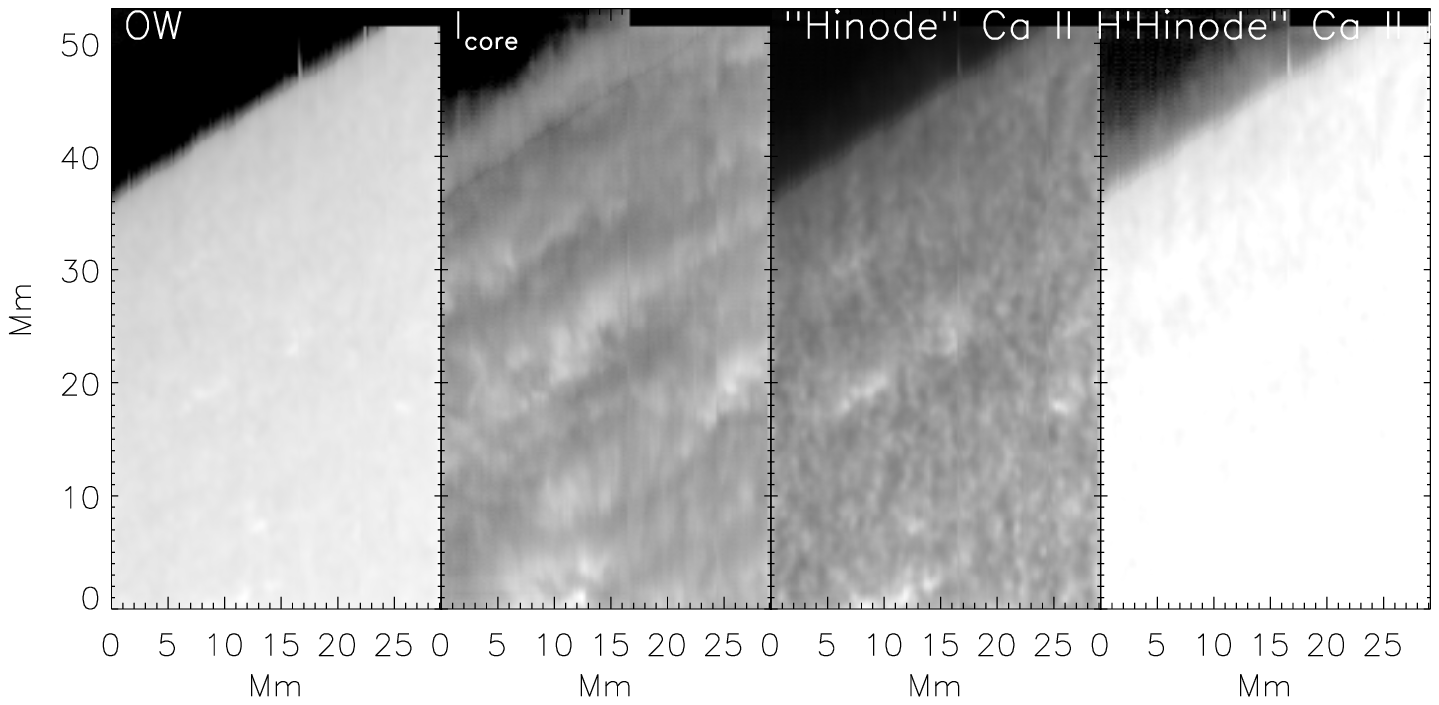}}\\$ $\\\\$ $\\
\centerline{\resizebox{4.4cm}{!}{\hspace*{1.cm}\includegraphics{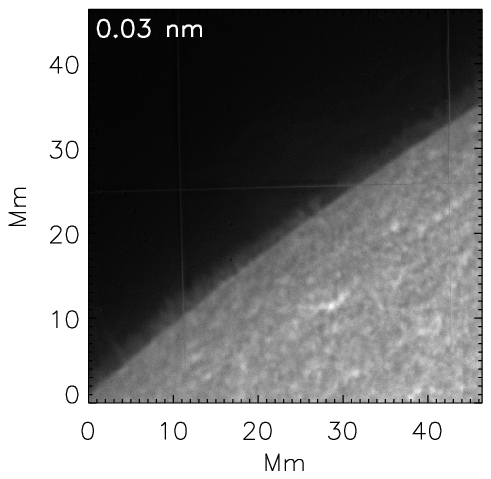}}\hspace*{-.5cm}\resizebox{4.4cm}{!}{\hspace*{1.cm}\includegraphics{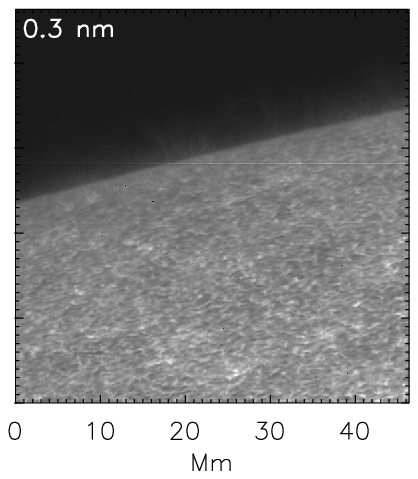}}}$ $\\$ $\\
\caption{Overview of the observations at the solar limb. {\em Top panel,
    left to right}: 2D maps obtained from the spectra in the outer wing (OW), the line core ($I_{\rm core}$, linear display), the synthetic Ca broad-band imaging data (linear display), and the same in logarithmic and clipped display.  {\em Bottom panel, left to right}: contrast-enhanced \ion{Ca}{ii} K Lyot-filter and Hinode \ion{Ca}{ii} H broad-band filter images at the solar limb. \label{limb_fov}}
\end{figure}
\begin{figure}
\resizebox{8.8cm}{!}{\includegraphics{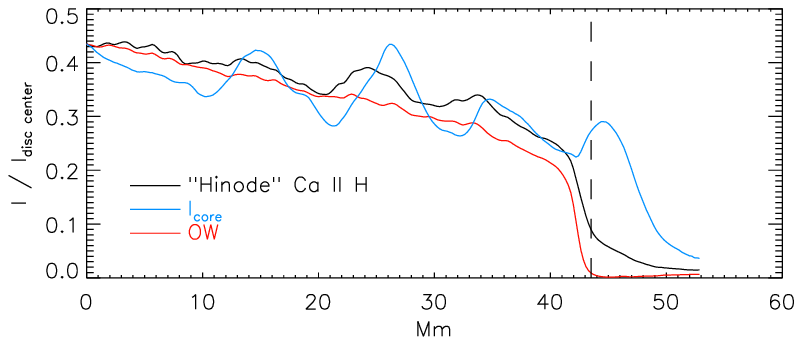}}\vspace*{-.5cm}
\resizebox{8.8cm}{!}{\includegraphics{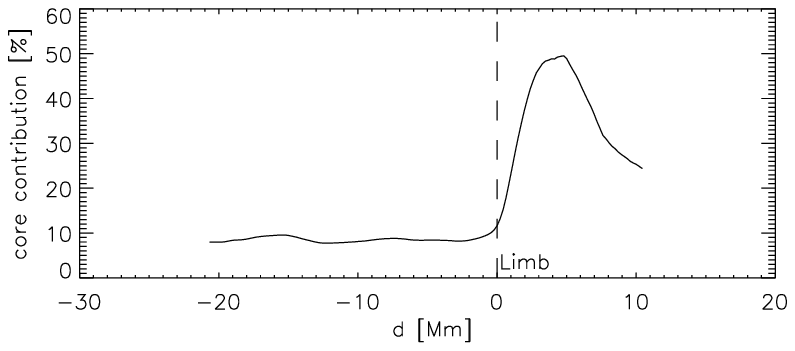}}
\caption{Intensity and relative core contribution on radial cuts 
    normalized to the intensity at disc centre. {\em Top panel}: average intensity in OW ({\em red line}), Ca line core $I_{\rm
      core}$ ({\em blue line}) and synthetic Ca broad-band imaging data ({\em
      black line}).  {\em Bottom panel}: relative contribution of the line core to the broad-band filter intensity. The {\em dashed vertical lines} denote
    the location of the limb.\label{fig_cuts}}
\end{figure}

As additional reference apart from the observed spectra we used \ion{Ca}{ii}
H profiles synthesized from the solar atmospheric models  FALC and FALF
\citep{fontenla+etal2006} in non-local thermal equilibrium (NLTE) with the RH
code \citep{uitenbroek2000}. \citet[][]{beck+etal2013a} discuss
the match of such NLTE profiles to individual and average observed Ca
spectra. These synthetic NLTE profiles were only multiplied with the
transmission profile of the 0.3\,nm filter ({\em lower panel} of
Fig.~\ref{filter}). For both synthetic profiles, the maximum of the
transmitted intensity is located near the Ca line core, but the spectral
extent of the emission peaks is small in comparison to the wavelength range in
the line wing that shows a transmitted intensity of comparable amplitude. 
To estimate the relative contributions of line wing and line core to the
  total intensity transmitted by the Hinode broad-band filter, we integrated
  the transmitted intensity from the outer wing up to 396.824\,nm (H$_{1V}$)
  for the former, and from 396.824 to 396.874\,nm for the latter. For the
  contribution from the red wing, which is not covered in the standard POLIS
  spectra, we assumed the same contribution as for the blue wing. On the disc,
the contribution of H$\epsilon$ at 397\,nm to the total intensity is negligible
(cf.~Appendix \ref{hepsi}). The
relative contribution of the Ca line core to the total intensity
transmitted through the filter is about 10\,\% for the FAL profiles.
\section{Results\label{results}}
\begin{figure*}
\begin{minipage}{17.6cm}
\begin{minipage}{8.8cm}
\resizebox{4.4cm}{!}{\hspace*{1.cm}\includegraphics{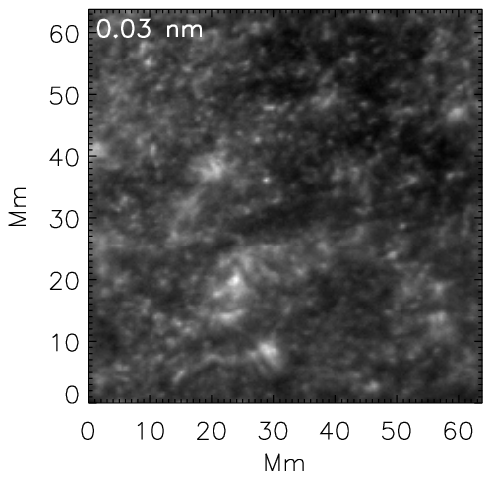}}
\resizebox{4.4cm}{!}{\hspace*{1.cm}\includegraphics{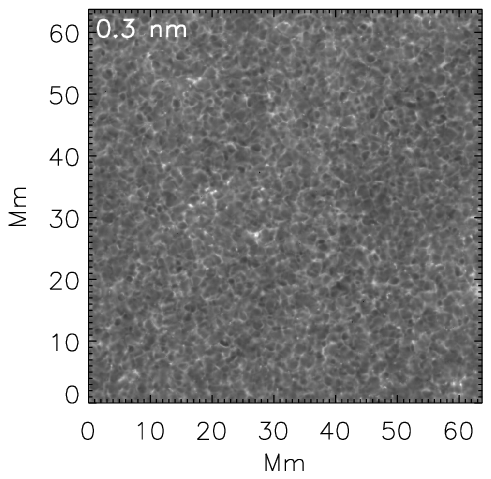}}
\end{minipage}\hspace*{.15cm}
\begin{minipage}{8.8cm}$ $\\
\resizebox{3.cm}{!}{\hspace*{1.cm}\includegraphics{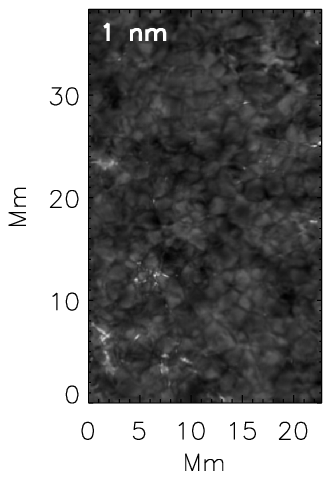}}\resizebox{3.cm}{!}{\hspace*{1.cm}\includegraphics{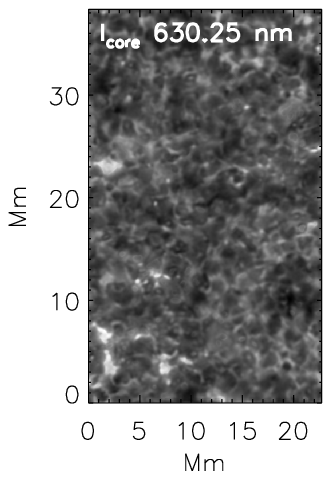}}\resizebox{3.cm}{!}{\hspace*{1.cm}\includegraphics{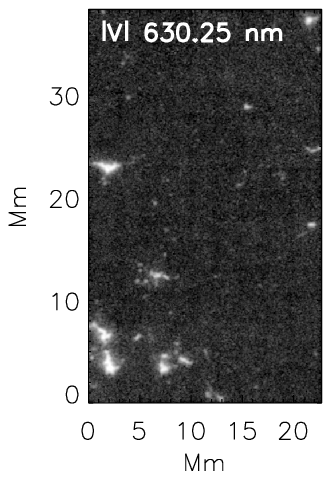}}
\end{minipage}$ $\\$ $\\$ $\\$ $\\
\resizebox{3.9cm}{!}{\hspace*{1.cm}\includegraphics{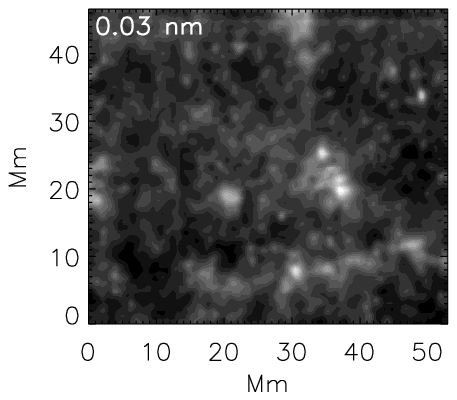}}\hspace*{.7cm}\resizebox{3.9cm}{!}{\hspace*{1.cm}\includegraphics{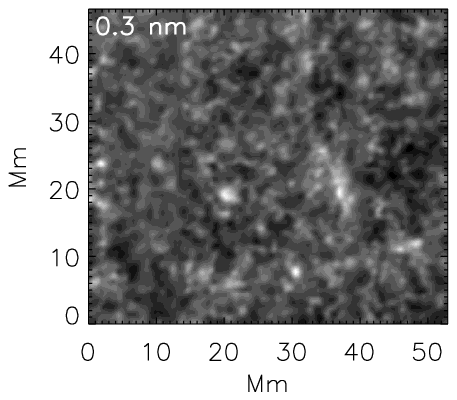}}\hspace*{1.cm}\resizebox{1.7cm}{!}{\hspace*{1.cm}\includegraphics{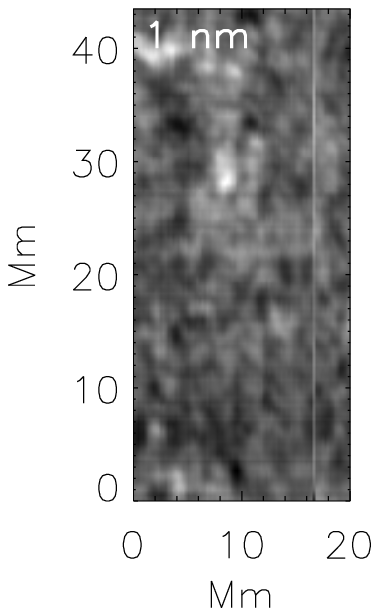}}$ $\\
\end{minipage}$ $\\
\caption{Disc centre observations with different filter widths. {\em Top row, left to right}: \ion{Ca}{ii} K Lyot image (0.03\,nm), Hinode/FG broad-band
  \ion{Ca}{ii} H image  (0.3\,nm), MFBD-reconstructed Ca image (1\,nm), co-spatial MOMFBD line-core image of \ion{Fe}{i} at 630.25\,nm
  and the unsigned wavelength-integrated Stokes $V$ signal of this line
  recorded with the GFPI. {\em Bottom row, left to right}: synthetic images
  derived by applying the corresponding filters of the top row to Ca spectra
  observed at disc centre. The synthetic image for the 1\,nm wide filter
    was obtained from the Ca spectra taken with the PCO camera at a different time. All images are scaled individually inside their full dynamical range.\label{ddisc}}
\end{figure*}
\subsection{Limb observations}
For the limb observations, the intensity in the outer wing \citep[OW, at
  about 396.397\,nm; cf.][]{rezaei+etal2007,beck+etal2008} of the Ca line and the line-core
intensity $I_{\rm core}$ of \ion{Ca}{ii} H (fixed wavelength at 396.85\,nm) are shown in the {\em top
    panel} of Fig.~\ref{limb_fov}, together with the synthetic Hinode
broad-band Ca imaging data. The latter map is shown twice with different display ranges (linear and logarithmic with clipping). Unlike in the Ca line-core image, the off-limb structures in the synthetic Ca imaging data cannot be seen in the linear display mode because of the strong intensity decrease beyond the limb (compare with Fig.~4 of DW12). The off-limb intensity in the line-core image is rather uniform with a minor variation along the limb.

The lower two panels of Fig.~\ref{limb_fov} show a \ion{Ca}{ii} K image near the limb taken with
  the 0.03\,nm-wide Lyot filter and a broad-band \ion{Ca}{ii} H image from Hinode for
  comparison. Both images were treated with an un-sharp masking to remove the
  radial intensity gradient and to enhance small-scale
  structures. Both filter images and the synthetic Ca broad-band image show a characteristic "grainy" structure of about 1\,Mm extent that is absent from the \ion{Ca}{ii} H line-core image. Note that the Ca line-core image  has exactly the same spatial resolution as the synthetic Ca broad-band image, therefore, the observed differences between the two images are real and can not be caused by any spatial resolution effects.

The {\em upper panel} of Fig.~\ref{fig_cuts} shows the average intensity in
the OW, the Ca line core and the synthetic Ca broad-band imaging data on radial cuts perpendicular to the limb. In the OW, the intensity drops to zero beyond the limb, whereas $I_{\rm core}$ shows a local maximum of intensity at about 2\,Mm
height above the limb. The limb location was determined visually in the OW map. The intensity of the synthetic Ca broad-band image ({\em black line}) decreases monotonically from about 10\,\% of disc-centre intensity right at the limb to 2\,\% at a height of 7\,Mm above the limb. Thus, the residual intensity in the synthetic Ca broad-band imaging data is only about 5\,\% of the intensity at disc centre for the height range that can be attributed to off-limb spicules.

The {\em lower panel} of Fig.~\ref{fig_cuts} shows that off the limb the
line-core region (396.824 to 396.874\,nm) contributes dominantly to the total synthetic Hinode broad-band filter intensity. The line-core
  contribution to the filter reaches only a maximum of about 50\,\% because
  the Ca line width off the limb exceeds the 50\,pm range attributed to the
  Ca line core here \citep[cf.~Fig.~5 of][]{beck+rezaei2011}. However, the relative contribution of the line core drops rapidly near the limb and remains at about 10\,\% on the disc.

\subsection{Disc centre observations \label{sec_obs}}
Figure \ref{ddisc} compares a set of Ca images taken with different filter
widths (0.03\,nm, 0.3\,nm and 1\,nm) at disc centre. The spatial
structuring changes significantly from one image to the next. The Lyot-filter
image shows spatially extended areas of low intensity of up to
10$^{\prime\prime}$ diameter \citep[cf.][]{rezaei+etal2008}, and an enhanced
intensity around the network. Granular structure is absent. The Hinode Ca
image is spatially much more uniform and exhibits an inverse granulation
pattern, on which a few isolated small-scale brightenings are overlayed. The
image taken with the broadest of the Ca filters (1\,nm) shows similar isolated
small-scale brightenings, but the global pattern has changed from inverse to
regular granulation instead. A comparison of the latter image with the
simultaneous, co-spatial line-core image and the unsigned
wavelength-integrated Stokes $V$ signal, $\int |V(\lambda)| d\lambda$, of the
\ion{Fe}{i} line at 630.25\,nm line (denoted by ``Stokes $|V|$'' in the following) extracted from the GFPI spectra ({\em
  rightmost two panels} in {\em top row} of Fig.~\ref{ddisc}) reveals the
strong imprint of photospheric magnetic fields on both the line-core
intensity image \citep[cf.][]{puschmann+etal2007} and the Ca imaging
data for a filter with a FWHM of 1\,nm, while the \ion{Fe}{i} line-core image
shows a background pattern of inverse granulation again. The bright features
in the 1-nm Ca filter image are significantly smaller than in the line-core
image of the photospheric line and show no intensity enhancements in
their surroundings, while both images have a comparable spatial resolution \citep{puschmann+beck2011}. 
\begin{figure}
\centerline{\includegraphics{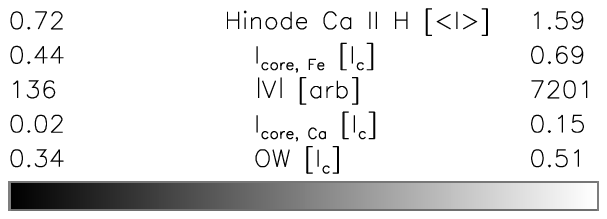}}$ $\\
\centerline{\resizebox{8.8cm}{!}{\hspace*{.75cm}\includegraphics{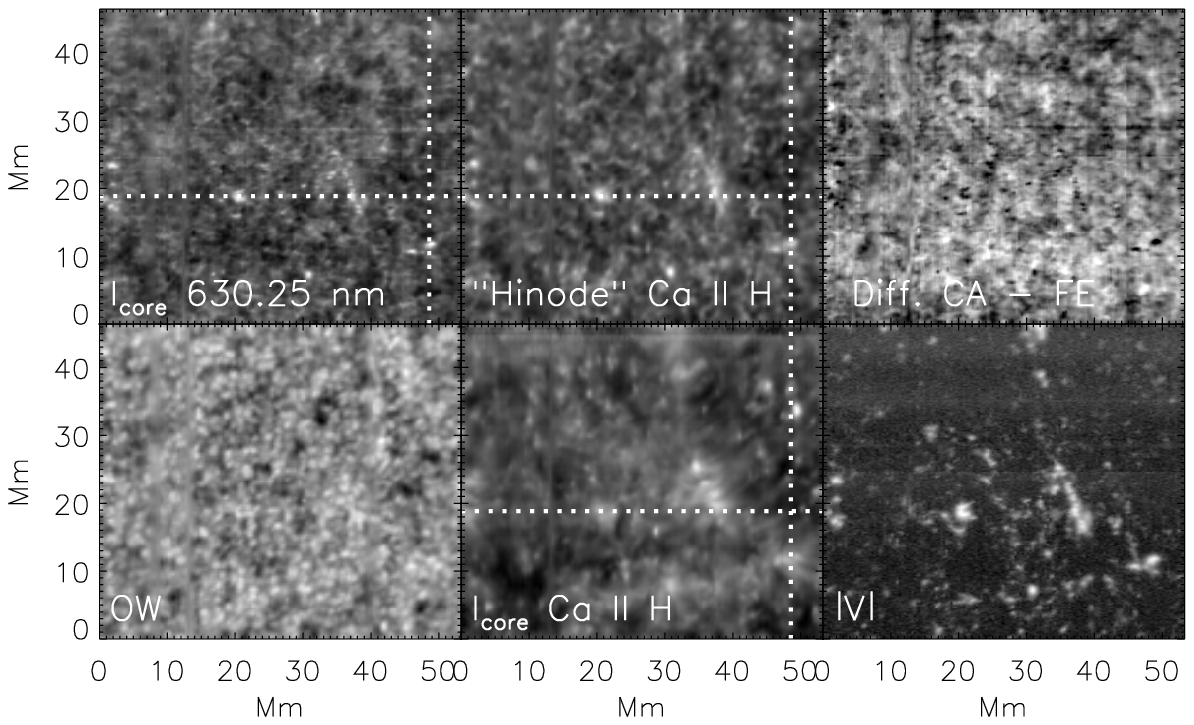}}}$ $\\
\caption{Overview maps of the POLIS observation at disc centre. {\em Bottom
    row, left to right}: OW, $I_{\rm core}$ of \ion{Ca}{ii} H and Stokes
    $|V|$ signal of the \ion{Fe}{i} line 630.25\,nm. {\em Top row, left to
      right}: $I_{\rm core}$ of \ion{Fe}{i} at 630.25\,nm, synthetic Ca
    broad-band imaging data and difference image of the latter two. The
      dotted white lines denote the locations of the cuts shown in Fig.~\ref{haze_cuts}.\label{center_fov}}
\end{figure}
\begin{figure}
\centerline{\resizebox{8.8cm}{!}{\includegraphics{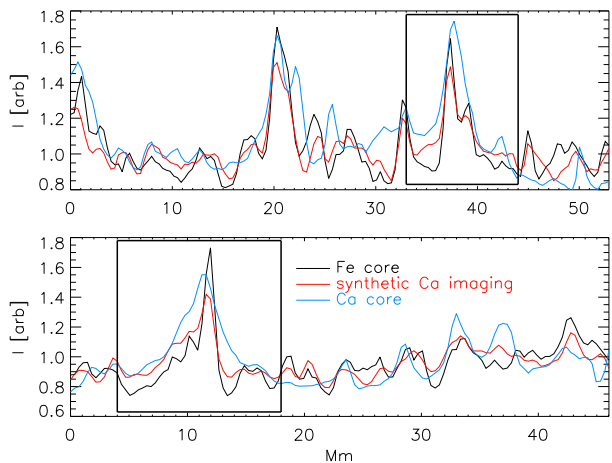}}}$ $\\
\caption{Cuts along the dotted white lines in Fig.~\ref{center_fov}
  through the Fe line-core map (black lines), the synthetic Ca broad-band
  imaging (red lines) and the Ca line-core intensity map (blue lines). All
  intensities have been re-scaled to the same dynamical range by linear
  regression. The black rectangles mark locations of network and their surroundings.\label{haze_cuts}}
\end{figure}

The maps in the {\em lower row} show the corresponding images obtained by
applying the respective filters of the upper row to observed Ca spectra. The change of the
background pattern from the absence of granulation through inverse granulation
towards regular granulation is well reproduced by the synthetic images. The
synthetic Lyot-type and 0.3\,nm filter images were obtained from the same set
of spectra and therefore have exactly the same spatial resolution. The extent
of the enhanced intensity (``haze'') at and around the magnetic
photospheric network (cf.~with the Stokes $V$ map in Fig.~\ref{center_fov})
reduces significantly for the 0.3\,nm-wide filter and is absent from the 1-nm
filter images.

Figure \ref{center_fov} shows maps of line parameters derived from the Ca
  spectra observed with POLIS at disc centre, the synthetic broad-band image,
  Stokes $|V|$ and the line-core intensity of the \ion{Fe}{i} line at
  630.25\,nm observed with the second channel of POLIS, and the difference image of the synthetic Ca broad-band
  and the \ion{Fe}{i} line-core map (compare with DW12, his Fig.~1, and
  RE09, their Fig.~7). For the difference image, the relative
  intensities of the Fe line-core map in the full field of view (FOV) were
  re-scaled by a linear regression to the intensity range of the synthetic Ca
  broad-band imaging data. The difference image highlights regions with
    a relative intensity excess in the broad-band Ca imaging data over the Fe
    line-core map. To visualize the haze around the network, we laid cuts
    through the Fe line-core map, the synthetic Ca broad-band image and the Ca
  line-core map (Fig.~\ref{haze_cuts}) after balancing their dynamical range
  in the same way by linear regression. The haze shows up as a weak
  intensity enhancement of the synthetic Ca broad-band image over the Fe line-core
  intensity. The Ca line-core image usually exceeds both of the other
  quantities over an even larger area.
\begin{figure}
\centerline{\includegraphics{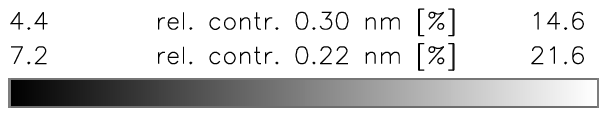}}$ $\\
\resizebox{8.8cm}{!}{\hspace*{.75cm}\includegraphics{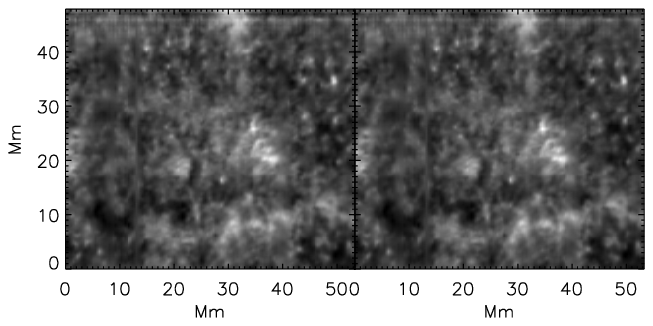}}$ $\\$ $\\
\caption{Relative contribution of the line core to the filter intensity at
  the disc centre. {\em Left/right}: for filters of FWHM\,=\,0.22 and 0.3\,nm, respectively.\label{fig1a}}
\end{figure}

The synthetic Ca imaging data match well to the line-core intensity of the photospheric \ion{Fe}{i} line, as noted by DW12, but the Ca line-core image shows a quite different pattern. Starting from the locations of the photospheric magnetic flux as outlined in the map of
$|V|$, elongated slender brightenings of several Mm length can be seen in the
Ca line-core intensity \citep[``fibrils'';][]{zirin1974,marsh1976,pietarila+etal2009,delacruzrodriguez+etal2011} that also have a counterpart in the line-of-sight (LOS) velocities of the Ca line core (cf.~BE09, their Fig.~5). 

Comparing the true Ca line-core image ({\em middle bottom panel} in
  Fig.~\ref{center_fov}, ``FWHM''\,=\,0.0038\,nm\footnote{Sampling limited
    spectral resolution.}) even with a (synthetic) Lyot-type filter image ({\em
    left column} in Fig.~\ref{ddisc}, FWHM 0.03\,nm) shows that the extent of
  the enhanced intensity around network fields is significantly larger and
  consequently the size of regions with reduced intensity is significantly
  smaller in the Ca line-core image. The difference image of the synthetic
  broad-band Ca imaging data and the Fe line-core map shows no traces of
  fibrils, unlike the Ca line-core image and the LOS velocity map. The comparison of the Ca line-core image with the Stokes $|V|$ signal
  reveals which of the Ca brightenings are not related to magnetic fields, e.g.~the prominent bright grain
  \citep[cf.][]{rutten+uitenbroek1991,carlsson+stein1997,beck+etal2013a} at
  $x, y \sim 50$\,Mm, $35$\,Mm. Such features are usually
  absent in the synthetic broad-band imaging data ({\em top middle panel} of Fig.~\ref{center_fov}), but appear in the synthetic Lyot-filter image ({\em lower-left panel} of Fig.~\ref{ddisc}). 

 For the average quiet Sun profile observed at disc centre, about 9\,\%
 (FALC: 10\,\%, FALF: 13\,\%) of the total intensity transmitted by the
 broad-band filter comes from the line-core  region (396.824 to
 396.874\,nm). The relative contribution of the line-core region to the filter
 intensity varies between 4 and 15\,\% across the observed FOV at disc centre
 (Fig.~\ref{fig1a}), with an average contribution of 8\,$\pm$\,1\,\%.  For
   the synthetic filter with 0.22\,nm FWHM, the corresponding numbers are
   between 7 and 21\,\%, with a mean value of 13$\pm1$\,\%. The largest
 values are attained in the centre of the magnetic network (compare
   Fig.~\ref{fig1a} with the unsigned Stokes $V$ map in
   Fig.~\ref{center_fov}), whereas the lowest values correspond to
 reversal-free profiles \citep[cf.][]{rezaei+etal2008}. The enhanced relative contribution of the line core to the total filter intensity on locations of photospheric magnetic fields does, however, not instantly imply a stronger chromospheric contribution to the broad-band filter imaging in the network. On locations of photospheric magnetic fields, the intensity of the Ca line core is raised at all wavelengths, i.e.~also outside of the emission core, by a contribution from an atmosphere with a locally increased temperature, but not necessarily a chromospheric temperature rise \citep[e.g.~Fig.~17 of][]{beck+etal2008}.
\begin{figure}
\centerline{\includegraphics{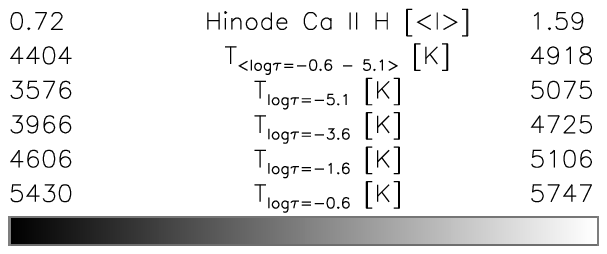}}$ $\\
\resizebox{8.8cm}{!}{\hspace*{.75cm}\includegraphics{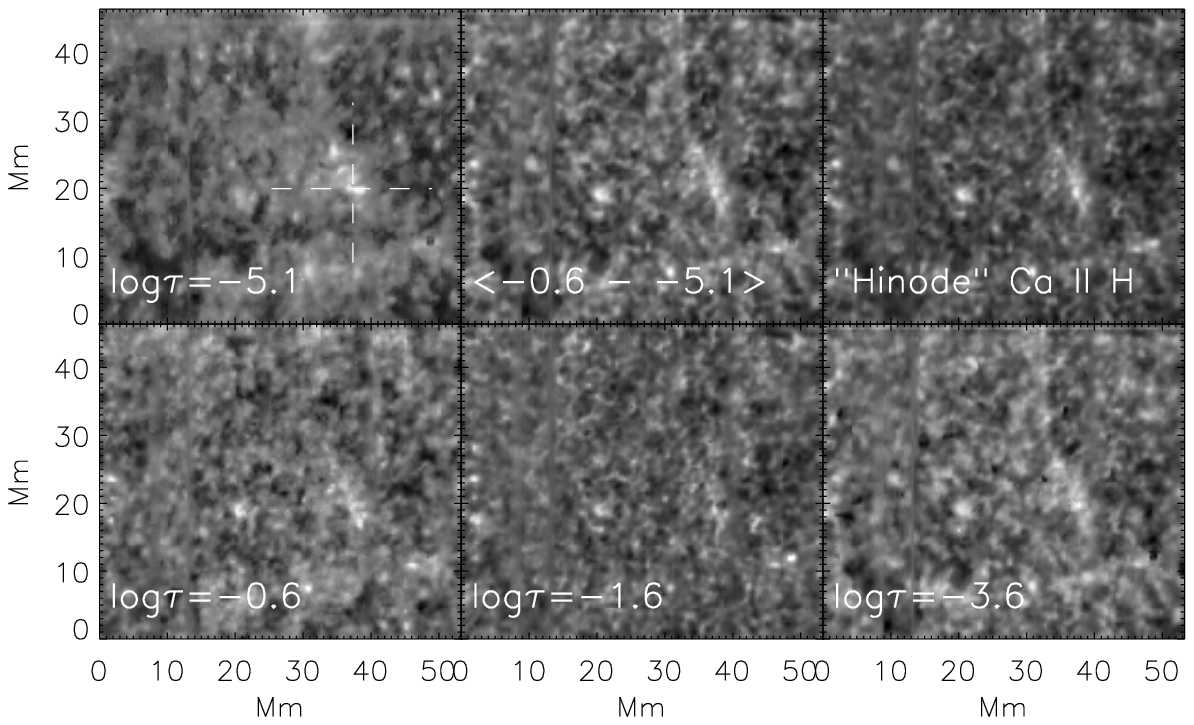}}\\$ $\\
\caption{Temperature maps at disc centre. {\em Bottom row, left to right: at} optical depths of $\log\tau = -0.6, -1.6,$ and $- 3.6$. {\em Top row, left to right}:  at $\log\tau = -5.1$, and averaged over $-0.6$ to $-5.1$. The {\em top rightmost panel} shows the synthetic Ca broad-band image. The {\em
    white dashed lines} in the temperature map at $\log\tau = -5.1$ denote the
  location of the spatial cuts shown in Fig.~\ref{last_fig}.\label{fig_center_t}}
\end{figure}

Figure \ref{fig_center_t} shows temperature maps at different
  atmospheric layers from the photosphere to the lower chromosphere. The
  temperature was obtained on an optical depth scale related to
$\tau_{500\,\rm nm}$ by applying the inversion approach of
\citet{beck+etal2013} to the observed Ca spectra that assumes local thermal
equilibrium (LTE). Comparing the synthetic Ca imaging data with the
temperature maps reveals a close match of the former to the temperature at
$\log \tau = -1.6$, and an acceptable match to the temperature averaged
between $\log \tau = -0.6$ and $-5.1$, but similar to the Ca line-core
intensity, the temperature map at $\log \tau = -5.1$ shows a different
structuring than the synthetic Ca broad-band imaging data.  Non-LTE
  effects should become important in the layers above $z\sim 400$\,km ($\log
  \tau \sim -3.2$) \citep[e.g.][]{rammacher+ulmschneider1992}, with the main
  effect of a decoupling between kinetic temperature and emergent
  intensity. If included in an inversion, NLTE effects would presumably
  increase the spatial temperature variations relative to the LTE case and
  therefore further reduce the similarity between the temperature maps above
  $\log \tau \sim -3.2$ and the synthetic Ca imaging data. The visual
comparison thus suggests that the intensity in the synthetic Ca imaging data originates from below the formation height of the line core, or $\log\tau = -5.1$, respectively, and pertains to a region that shows no clear signature of fibrils in either intensity or velocity maps. 
\begin{figure}
\centerline{\hspace*{.5cm}\includegraphics{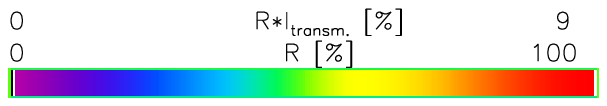}}$ $\\
\resizebox{8.8cm}{!}{\hspace*{1.cm}\includegraphics{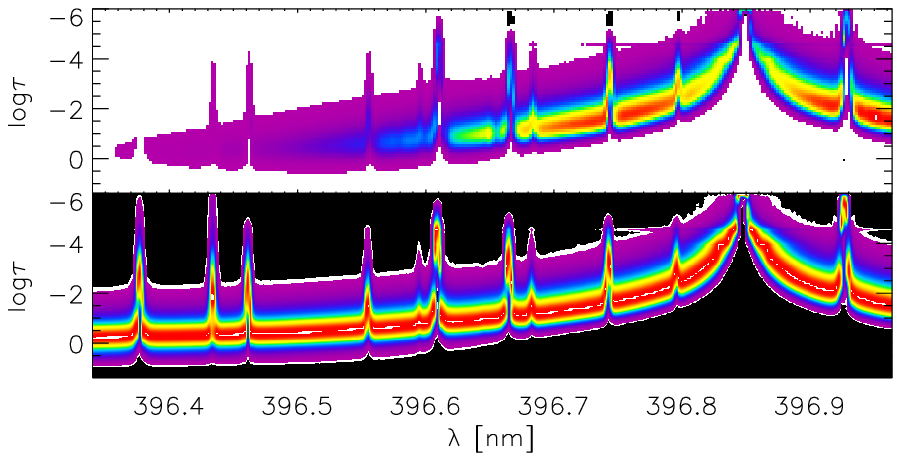}}\\$ $\\
\caption{Intensity response functions. {\em Bottom}: intensity response of individual wavelengths in \ion{Ca}{ii} H to different layers of optical depth. {\em Top}: the former multiplied with the intensity spectrum transmitted by a broad-band filter (cf.~Fig.~\ref{filter}).\label{contrib}}
\end{figure}

\begin{figure}
\resizebox{8.8cm}{!}{\includegraphics{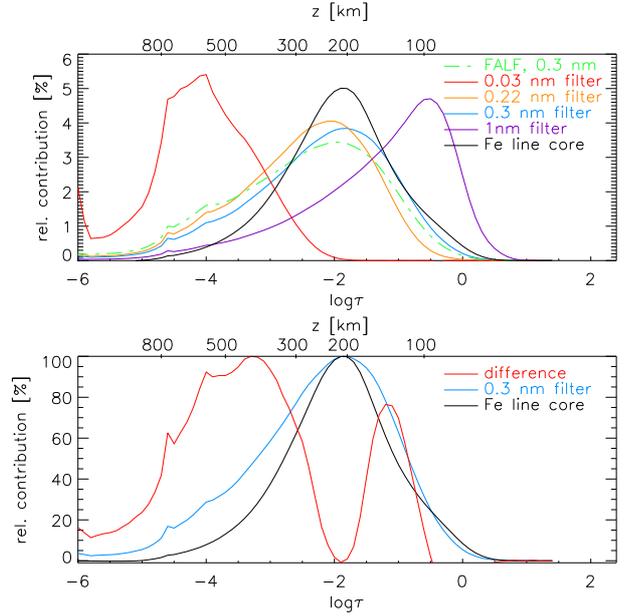}}
\caption{Relative intensity response functions. {\em Top panel}:
    intensity response in synthetic Ca imaging data for filters with FWHM of
    0.03\,nm, 0.22\,nm, 0.3\,nm, and 1\,nm, respectively ({\em red/orange/blue/purple lines}). {\em Green-dash-dotted line}: response for FALF profile passed through a 0.3\,nm filter {\em Black line:} intensity response of the 630.25\,nm line core. {\em Bottom panel}: response for a 0.3\,nm filter and the 630.25\,nm line core normalized to maximal response ({\em blue/black line}), and their difference ({\em red line}).  \label{contrib1}}
\end{figure}

To obtain a quantitative measure of the formation heights that contribute to the synthetic Ca broad-band imaging data, we calculated the intensity response function \citep[cf.][]{cabrera+bellot+iniesta2005} $R$ of wavelengths in the \ion{Ca}{ii} H line in LTE \citep[{\em lower  panel} of Fig.~\ref{contrib}; see also][]{carlsson+etal2007,pietarila+etal2009,jafarzadeh+etal2013}. The values at each wavelength $\lambda$ were separately normalized to the maximum response $R(\lambda)$, therefore the response function holds individually for each wavelength. The
filter imaging performs, however, an integration in wavelength over an
additionally wavelength-dependent intensity $I(\lambda)$, which yields an
unequal contribution of different wavelengths, i.e.~a contribution weighted
with both $I(\lambda)$ and the pre-filter curve (cf.~Fig.~\ref{filter}). The
{\em upper panel} of Fig.~\ref{contrib} shows the relative intensity response
when an average quiet Sun Ca profile is transmitted through a broad-band
filter, i.e.~a multiplication of the response function at a given wavelength
$\lambda$  by the transmitted intensity $I(\lambda)$. The transmitted intensity in the line wing is higher than in the line core, which causes a stronger relative contribution of optical depth layers between $\log\tau
= -1$ and $-4$ than for higher layers ($\log\tau < -4$). The wavelength-integration executed by the broad-band filter
imaging then corresponds to an averaging of the intensity-weighted response
function over the filter extent. This yields finally an intensity response
function with optical depth, but without wavelength dependence (Fig.~\ref{contrib1}). 

We used the tabulated relation between optical depth and geometrical
  height of the Harvard Smithsonian Reference Atmosphere
  \citep{gingerich+etal1971} to provide approximate formation heights in
  absolute geometrical units opposite to a derivation of the height scale
    from the inversion results themselves as done in \citet{puschmann+etal2005,puschmann+etal2010}. The response function for the 1\,nm-wide filter
  peaks at about 100\,km, with a monotonically decreasing contribution from
  atmospheric layers above. The two broad-band Ca filters with 0.3\,nm and
  0.22\,nm peak at about $\log \tau= -2$ ($z\sim200\,$km, similar to  the
  247\,km obtained by \citeauthor{carlsson+etal2007} \citeyear{carlsson+etal2007}) and have only small
  contributions from layers above  $\log \tau= -3$. Their relative
  contributions reduce again monotonically with height in the atmosphere. The
  difference between transmitting the average observed Ca profile ({\em blue
    line}) or the synthetic FALF profile ({\em green dash-dotted line}) through a
  0.3\,nm filter is minor, even if the latter has a slightly larger 
  contribution above $\log\,\tau < -3$. A Lyot-type narrow-band filter, on the
  other hand, has significant contributions of layers up to 800\,km, although
  the contribution peaks at about 600\,km. The very line core of \ion{Ca}{ii} H
  (line not drawn) would peak at the upper end of the optical depth scale,
  i.e.~$\log\,\tau = -6$ ($z\sim 1500\,$km), in the LTE approximation. The
  line core of the \ion{Fe}{i} line at 630.25\,nm behaves similar to, e.g.~the
  0.3\,nm filter, but has less contributions from upper atmospheric layers
  than this filter (Fig.~\ref{contrib1}). The contribution curves compare well with the visual appearance in Figs.~\ref{ddisc} and \ref{center_fov}: granular background pattern for the 1\,nm-wide filter, inverse granulation at 0.22 to 0.3\,nm, and absence of granulation for 0.03\,nm. The relative contribution from heights above 800\,km are 13, 4, 3 and about 1\,\%, for a 0.03, 0.22, 0.3 and 1\,nm wide filter, respectively.

To determine where the difference intensity between the Fe line-core image and a 0.3\,nm-wide filter should mainly come from, we normalized these two response functions to their maximum value instead of the total area. This allows one to roughly quantify the contribution height range of the difference image ({\em lower panel} of Fig.~\ref{contrib1}) because both functions peak at about the same height. The difference curve shows two local maxima at about 150\,km and 400\,--\,500\,km, respectively. About 11\,\% of the difference originate from layers above 800\,km, 67\,\% from between 200 to 800\,km and 16\,\% from heights below 200\,km. Thus both the response of a 0.3\,nm-wide filter, and the difference image between such a filter and a Fe line-core image have at maximum about 10\,\% contribution from atmospheric layers above 800\,km.
\section{Summary and discussion\label{discus}}
The interpretation of the broad-band (FWHM of 0.3\,nm) \ion{Ca}{ii} H filter
imaging data of the Hinode Solar Optical Telescope (SOT) as a
{``chromospheric'' measure, as done for example in
    \citet{katsukawa+etal2007}, \citet{mitrakaev+etal2008},
    \citet{perez+etal2008}, \citet{guglielmino+etal2008},
    \citet{socasnavarro+etal2009}, \citet{yurchyshyn+etal2010},
    \citet{park+chae2012} or \citet{gupta+etal2013}, requires that
  chromospheric layers dominate the total wavelength-integrated filter
  intensity. Whereas beyond the solar limb this is ensured, the situation for
  observations on the solar disc is quite different. \citet{mcintosh+depontieu2009} and \citet{depontieu+etal2009} used a
    high-pass filter on Hinode \ion{Ca}{ii} H imaging data, excluding
    frequencies below 18\,mHz (periods $>$ 60\,secs) to retain
    only what they called ``upper chromospheric activity''. Characteristic
    intensity de-correlation times reach, however, 60\,secs already at
    heights of about 400\,km \citep[cf.~Fig.~13 in][]{beck+etal2008}, so the
    effectiveness of such a high-pass filtering is not instantly
    clear. \citet{reardon+etal2009} found no good resemblance between
    broad-band Ca imaging and the line core of \ion{Ca}{ii} IR spectra neither
    on a temporal average nor in their temporal evolution considering frequencies up to 15 mHz.

We find that the off-limb intensity in broad-band imaging data that can be
attributed to upper chromospheric features such as spicules seen at heights of
about 1\,Mm or more above the limb is only about 5\,\% of the intensity on the
disc (Fig.~\ref{fig_cuts}). The off-limb intensity stems exclusively from
emission in the line-core region of \ion{Ca}{ii} H \citep[cf.~the average
off-limb profiles in][]{beck+rezaei2011} and H$\epsilon$ (cf.~Appendix \ref{hepsi}) and cannot contain any contribution
from photospheric radiation. It is, however, the result of an integration
along a LOS roughly parallel to the solar limb, hence the solar surface,
i.e.~it adds up emission over an extended spatial
distance at a roughly constant height in the atmosphere. The integrated
intensity in the Ca line core for a LOS that is perpendicular to the solar
surface will be significantly lower, but it requires a detailed calculation
such as in \citet{judge+carlsson2010} to determine accurately the amount. The
investigation of the relative contribution of the H$\epsilon$ line at 397\,nm
to broad-band imaging  shows that its line depth in
absorption on the disc for a vertical LOS is significantly lower than its
amplitude of emission for a horizontal LOS. The total off-limb emission of
5\,\% in the synthetic filter image thus is an upper limit for its
contribution to a LOS at disc centre.

On the disc, on average only about 10\,\% of the intensity that a broad-band
filter transmits correspond to the line-core region (Fig.~\ref{filter}), which
covers formation heights from the temperature minimum upwards. The maximal
contribution of the line-core region of about 15\,\% is attained at the centre
of the magnetic network. Together with the off-limb filter
  intensity, this puts  an upper limit of 15\,\% as maximal chromospheric, hence spicular contribution to broad-band Ca filter imaging data at disc centre. The true spicular contribution to the broad-band filter imaging data could be quite overestimated by this. On the one hand, the complete Ca line-core region shows an increased intensity on magnetic locations -- that are implicitly taken as locations of spicules here -- which implies an increased temperature throughout all of the atmosphere, i.e.~also at photospheric layers \citep[][]{beck+etal2008}. On the other hand, the Ca line core at disc centre shows an absorption profile with overlayed emission, so it cannot only be caused by spicular emission. Since the spicules exhibit emission profiles with about 5\,\% of the intensity at the centre of the disc and the relative contribution  of the line-core region to the total filter intensity to the filter is 15\,\% at maximum, the effective contribution of spicules to a broad-band filter image could in reality be below 1\,\% at disc centre. A similar estimate is provided by the intensity response function, where only about 3\,\% of the relative contribution come from heights above 800\,km for a 0.3\,nm-wide filter.   

A comparison with Ca line-core images, synthetic maps for various filter
widths, and temperature maps at different optical depth layers reveals that
synthetic Ca broad-band imaging data match the line-core intensity of
the photospheric Zeeman-sensitive \ion{Fe}{i} line at 630.25\,nm and the
temperature at layers (far) below $\log \tau = -4$. The synthetic Ca
broad-band imaging data lack the traces of fibrillar structures that appear in the Ca line-core intensity and LOS velocity above $\log \tau =- 4$. Even a Lyot-type filter image shows at most weak indications of fibrillar structures. Imaging data with a 1\,nm wide Ca filter provide only information on solar atmospheric layers below the formation height of the line cores of photospheric lines  (Figs.~\ref{ddisc} and \ref{contrib1}). It would thus be better to use some Lyot-type filter in the blue imaging channel of the GREGOR Fabry-P{\'e}rot Interferometer \citep{puschmann+etal2012a,puschmann+etal2012c,puschmann+etal2012b} instead of such a broad filter in future observations. This would allow nearly chromospheric context imaging in addition to the photospheric spectropolarimetric data taken with the GFPI. The future integration of the Blue Imaging Solar Spectrometer \citep[BLISS,][]{puschmann+etal2012c,puschmann+etal2013} will then allow also the parallel recording of \ion{Ca}{ii} H spectra at the 1.5 meter GREGOR telescope \citep{schmidt+etal2012a}. 

The intensity response function weighted by an average quiet Sun Ca profile transmitted through a broad-band Ca filter locates the main contributions at about $\log\tau = -2$ ($z\sim200\,$km). The relative contribution reduces monotonically with height. A quantitative estimate of the formation height for the difference image between the line-core intensity of the \ion{Fe}{i} line at 630.25\,nm and broad-band Ca filter imaging has the strongest contributions between 200 and 800\,km, and roughly equally strong contributions from above 800\,km and below 200\,km. 

With a maximal contribution of about 10\,\% for layers above 800\,km, we
therefore suggest that for observations at disc centre the broad-band Hinode
Ca imaging data trace mainly upper photospheric layers ($z\ll 600$\,km)
dominated by reverse granulation (cf.~Figs.~\ref{ddisc} and \ref{center_fov}),
but not upper chromospheric structures ($z > 1$\,Mm), as was also concluded by
RE09. A chromospheric contribution ($z>800\,$km) of up to 10\,\% still does
not allow one to use such data for studying the chromosphere at disc centre
because it is impossible to isolate the corresponding information from the other 90\,\% of contribution to the integrated intensity image.
\begin{figure}
\centerline{\hspace*{.5cm}\includegraphics{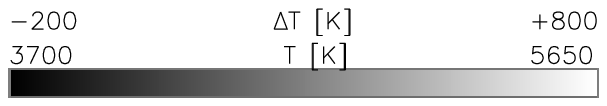}}$ $\\
\resizebox{8.8cm}{!}{\hspace*{1cm}\includegraphics{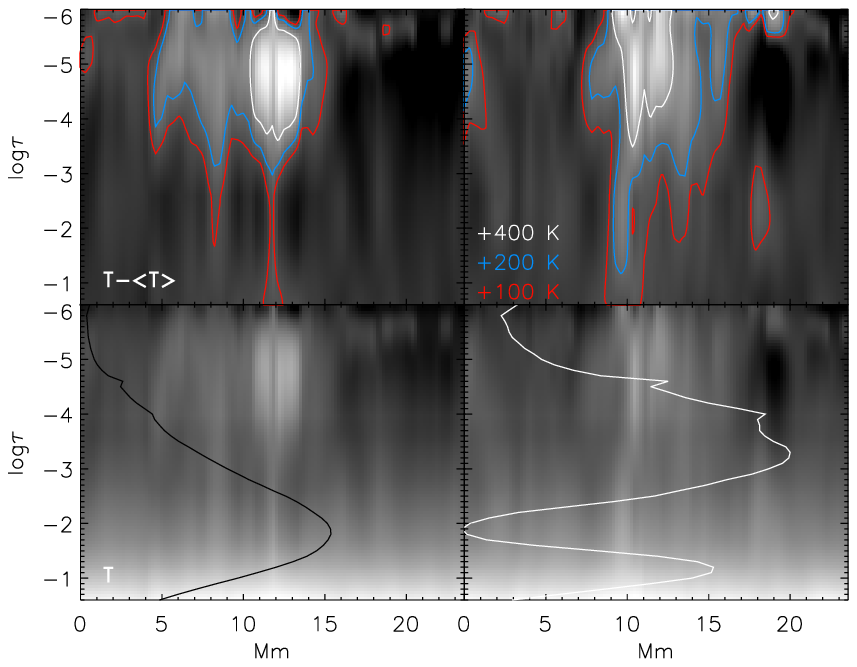}}\\$ $\\
\caption{Spatial cuts through temperature maps at disc centre. {\em Bottom
    row}: modulus of the temperature. {\em Top row}: relative temperature
  fluctuations after subtraction of the average temperature
  stratification. {\em Left column}: cut in $y$. {\em Right column}: cut in
  $x$. The {\em black} and {\em white line} in the {\em bottom row}
    indicate the response of a 0.3\,nm-wide filter and the difference image of
    broad-band Ca imaging and \ion{Fe}{i} line-core intensity, respectively.\label{last_fig}}
\end{figure}

Now, if the identification of the faint haze around the photospheric
  magnetic network with type-II spicules in DW12 is thus strongly unlikely,
  the question remains what causes the haze instead. The results of the LTE
inversion of the \ion{Ca}{ii} H spectra provide a different
  possible explanation for the haze. Figure \ref{last_fig} shows the modulus
of the temperature and the relative temperature variations on two spatial cuts
across a network element \citep[see also Fig.~16 of][]{beck+etal2013a}. The location of the cuts is indicated by the {\em
  white dashed lines} in the temperature map at $\log\tau = -5.1$ of
Fig.~\ref{fig_center_t}. Especially in the relative temperature fluctuations,
a thermal ``canopy'' appears next to the photospheric flux concentrations
located at $x\sim 12$\,Mm in the {\em left panel} and at $x\sim 10$\,Mm in the
{\em   right panel}. These thermal canopies extend up to a few Mm away from
the location of the photospheric magnetic fields, and appear at layers between
$\log\tau \sim -3$ up to $-6$, which corresponds to heights between
500\,--\,600\,km up to 2\,Mm. Similar thermal canopies were recently
  identified by \citet{delacruzrodriguez+etal2013} in numerical simulations
  and observations of \ion{Ca}{ii} IR spectra. With the location of the
maximum intensity response below $\log \tau = -4$ ($z\sim 600$\,km, {\em
    bottom left panel} of Fig.~\ref{last_fig}), the Hinode Ca imaging data
are thus presumably sampling the lower end of these thermal canopies,
which yield the haze around the photospheric magnetic network in the imaging
data. The intensity response of the difference image of broad-band Ca
  imaging and photospheric \ion{Fe}{i} line-core intensity peaks at the lower boundary of the thermal canopies ({\em bottom right} panel of Fig.~\ref{last_fig}).
\section{Conclusions\label{schluss}}
The broad-band (FWHM $\sim$0.3\,nm) Ca filter imaging data of the Hinode SOT are dominated by upper photospheric layers ($z < 600$\,km) in observations at disc centre. The faint haze of increased intensity around the photospheric network should not be related to entirely chromospheric structures such as spicules. We suggest that the haze rather traces thermal canopies below a height of 1\,Mm that are related to concentrated photospheric magnetic flux. Any direct measurement of spicules with broad-band Ca filter imaging at disc centre should be impossible given the dominant intensity contribution (about 90\,\%) of the line wings, hence photospheric layers to the intensity transmitted by a broad-band interference filter.
\begin{acknowledgements}
The VTT is operated by the Kiepenheuer-Institut f\"ur Sonnenphysik (KIS) at the
Spanish Observatorio del Teide of the Instituto de Astrof\'{\i}sica de Canarias (IAC). The POLIS instrument has been a joint development of the High Altitude Observatory (Boulder, USA) and the KIS. C.B.~acknowledges partial support by the Spanish Ministry of Science and Innovation through project AYA2010--18029 and JCI-2009-04504. R.R. acknowledges financial support by the DFG grant RE 3282/1-1. We thank the referee for pointing out the importance of H$\epsilon$ to us.
\end{acknowledgements}
\bibliographystyle{aa}
\bibliography{references_luis_mod_partIII}
\begin{appendix}
\section{Contribution of H$\epsilon$ to the Hinode broad-band Ca filter\label{hepsi}}
\begin{figure}
\centerline{\resizebox{6.cm}{!}{\hspace*{1cm}\includegraphics{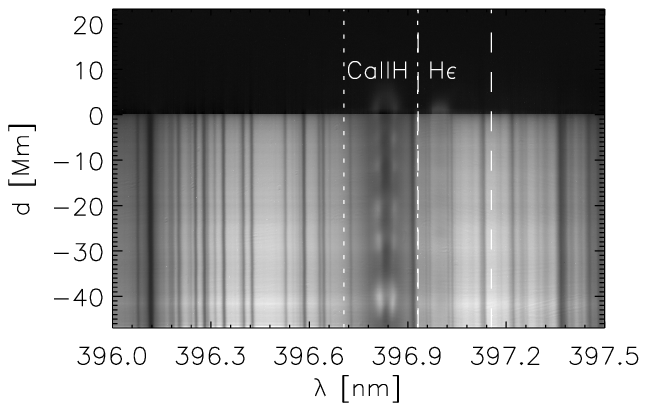}}}$ $\\
\resizebox{8.8cm}{!}{\includegraphics{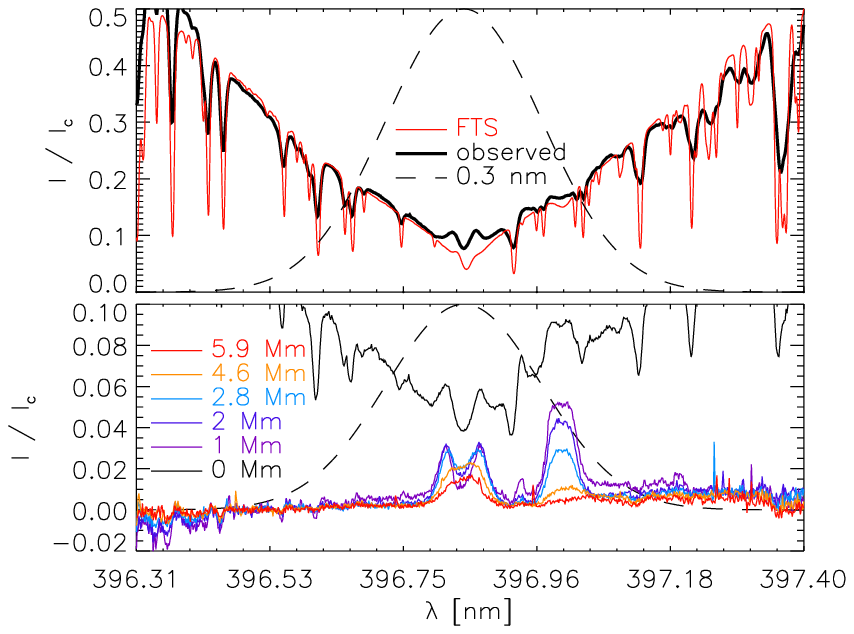}}\\$ $\\
\resizebox{8.8cm}{!}{\includegraphics{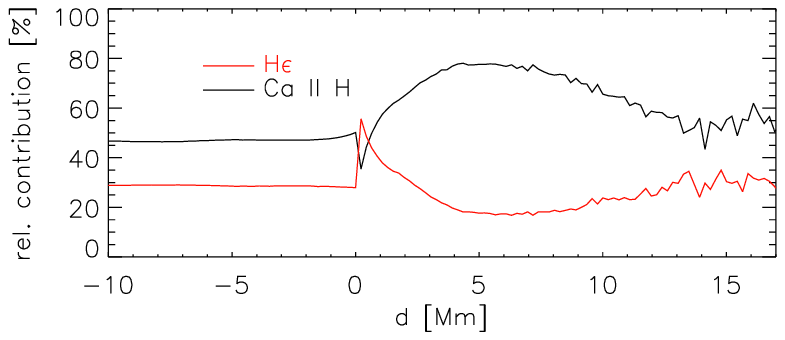}}
\caption{Relative contributions of \ion{Ca}{ii} H and H$\epsilon$ to the
  Hinode broad-band Ca filter. {\em Top panel}: average \ion{Ca}{ii} H spectra near the
  limb. The vertical dotted and dashed lines denote the wavelength ranges attributed to
  \ion{Ca}{ii} H and H$\epsilon$, respectively. {\em Second panel from the
    top}: average observed on-disk
  (thick black line) and FTS  spectrum (red line), and 0.3\,nm-wide prefilter curve (dashed
  line, scaled arbitrarily).  {\em Third panel from the top}: observed spectra at different limb
  distances. {\em Bottom panel}: relative contribution of \ion{Ca}{ii} H
  (black line) and H$\epsilon$ (red line) to a 0.3\,nm-wide prefilter.\label{fig_hepsi}}
\end{figure}
The 0.3\,nm-wide Hinode Ca prefilter was selected to record intensity
  images from wavelengths around the \ion{Ca}{ii} H line core. However, close to the
  \ion{Ca}{ii} H line core at 396.85\,nm, one can also find one of the Balmer
  lines, namely H$\epsilon$ at about 397\,nm. H$\epsilon$ is a chromospheric
  line like H$\alpha$, but shows only a small line depth on the disc (cf.~the
  FTS atlas spectra in the top panel of Fig.~\ref{fig_hepsi}). Here we
  investigate the amount that emission in H$\epsilon$ contributes to off-limb
  data taken with a broad-band Ca prefilter. 

The \ion{Ca}{ii} H spectra recorded with the PCO 4000 camera cover a
  larger spectral range than the default POLIS Ca CCD. Apart from the data set
on disc center used above, we also have some spectra near and beyond the limb
taken on 30 Jun 2010 available (top panel of Fig.~\ref{fig_hepsi}). The
stray-light correction for these data was done by subtracting a fraction of an
average off-limb profile \citep[cf.][]{marian+etal2012}, with the necessary
fraction being determined from the residual intensity in the blue line wing
far away from the chromospheric lines. The Hinode broad-band Ca prefilter
extends far enough into the Ca line wing to cover also the chromospheric
H$\epsilon$ line at 397\,nm (middle two panels of
Fig.~\ref{fig_hepsi}). Beyond the limb, the H$\epsilon$ line goes into
emission. The amplitude of the H$\epsilon$ emission exceeds that of
\ion{Ca}{ii} H in some height range above the limb ($d \sim 0$ to 3\,Mm; third
panel from the top in Fig.~\ref{fig_hepsi}). 

We estimated the relative contribution of \ion{Ca}{ii} H and H$\epsilon$ to
the total synthetic filter intensity for a 0.3\,nm-wide prefilter centred on
the \ion{Ca}{ii} H line core by calculating the fraction of intensity that
comes from the wavelength ranges marked in the top panel of
Fig.~\ref{fig_hepsi} (396.721 to 396.939\,nm and 396.939 to 397.156\,nm for
\ion{Ca}{ii} H and H$\epsilon$, respectively). The contribution of H$\epsilon$
slightly exceeds the one of \ion{Ca}{ii} H (bottom panel of
Fig.~\ref{fig_hepsi}) close to the limb ($d\sim 0$\,Mm; $> 50$\,\%
contribution). However, the stray-light correction included a step function
with zero correction on the disk and full correction beyond the -- manually
set -- limb position, so the results close to the assumed limb position ($|d|
< 0.3$\,Mm) might depend to some extent on the stray-light correction and the chosen limb location. Profiles at $d \sim 1$\,Mm should have
a clean and uncritical stray-light correction and therefore should be fully
reliable. These profiles still show a larger emission amplitude in H$\epsilon$
than in \ion{Ca}{ii} H (third panel from the top in Fig.~\ref{fig_hepsi}). The emission amplitude of H$\epsilon$ decreases faster
with increasing limb distance than the one of \ion{Ca}{ii} H, which exceeds the
former for $d> 3$\,Mm. The relative contribution of H$\epsilon$ to the Hinode
broad-band Ca filter decreases from about 50\,\% at the limb to about 20\,\%
at $d= 5$\,Mm and remains at this value for larger heights. For the  typical
height range of spicules (0 to 5--6\,Mm), the relative contribution of
H$\epsilon$ is about 1/3 of the total filter intensity. This could have a
visible and significant effect in a detailed modeling of the Hinode Ca imaging
data such as done in \citet{judge+carlsson2010}. The H$\epsilon$ line is
located in the wing of the filter transmission curve where its slope is steep,
which makes the amount of transmitted intensity sensitive also to the Doppler
shifts of H$\epsilon$, not only the amplitude of its emission. 
\end{appendix}

\end{document}